\documentclass[final,authoryear,3p,times]{elsarticle}
\usepackage{graphics}
\usepackage{amssymb}
\usepackage{hyperref}

\journal{Journal of Informetrics}

\begin{document}
\begin{frontmatter}

\title{Growth and structure of Slovenia's scientific collaboration network}

\author{Matja{\v z} Perc\corref{mp}}
\cortext[mp]{Electronic address: \href{mailto:matjaz.perc@uni-mb.si}{matjaz.perc@uni-mb.si}; Homepage: \href{http://www.matjazperc.com/}{http://www.matjazperc.com/}}
\address{Department of Physics, Faculty of Natural Sciences and Mathematics, University of Maribor,\\Koro{\v s}ka cesta 160, SI-2000 Maribor, Slovenia}

\begin{abstract}
We study the evolution of Slovenia's scientific collaboration network from $1960$ till present with a yearly resolution. For each year the network was constructed from publication records of Slovene scientists, whereby two were connected if, up to the given year inclusive, they have coauthored at least one paper together. Starting with no more than $30$ scientists with an average of $1.5$ collaborators in the year $1960$, the network to date consists of 7380 individuals that, on average, have $10.7$ collaborators. We show that, in spite of the broad myriad of research fields covered, the networks form ``small worlds" and that indeed the average path between any pair of scientists scales logarithmically with size after the largest component becomes large enough. Moreover, we show that the network growth is governed by near-liner preferential attachment, giving rise to a log-normal distribution of collaborators per author, and that the average starting year is roughly inversely proportional to the number of collaborators eventually acquired. Understandably, not all that became active early have till now gathered many collaborators. We also give results for the clustering coefficient and the diameter of the network over time, and compare our conclusions with those reported previously.
\end{abstract}

\begin{keyword} scientific collaboration \sep networks \sep preferential attachment \sep small worlds \sep Slovenia
\end{keyword}

\end{frontmatter}

\section{Introduction}

The structure of social networks is paramount for understanding the spread of knowledge, cultural traits, disease, as well as many other entities and attributes that can be associated with individuals living in groups or societies. As such it has been the subject of intense investigation, both theoretical as well as empirical, for at least half a century \citep{wassermanXbook94, wattsXbook99, barabasiXbook02, christakisXbook09}. Primarily, and in many ways not really surprisingly, however, these investigations were in the domain of social rather than natural sciences. Probably best know in this context is the study by \citet{milgramXptod67}, who studied how many steps it took, on average, to get a letter from a randomly selected person to a stockbroker in Boston, who was a friend of Milgram's. The result was six -- a number that has since been reused outside of science for a number of purposes, one of the latest examples being the launch of \href{http://www.sixdegrees.org/}{SixDegrees.org} seeking to exploit the ``small-world phenomenon" for charitable purposes. A shortcoming of the study of Milgram, as well as that of many others conducted in a similar fashion, is that the size and structure of social networks mapped in such a direct and labor intensive way is rather small and receptive to bias. The advent of large-scale online portals made it possible to test the ``six degrees of separation" hypothesis more thoroughly. Remarkably though, a study performed by \citet{leskovecXwww08}, encompassing some $30$ billion conversations from $240$ million people, reported that the average path length among Microsoft Messenger users is $6.6$. Although being closer to seven than six, the number is nevertheless in a strikingly good agrement with the result by Milgram obtained over 40 years ago.

In natural sciences the attention to networks was sparked by works such as those of \citet{wattsXnat98} and \citet{barabasiXsci99}, making ground-breaking advances with regard to our understanding of the ``small-world phenomenon" and the emergence of scaling via growth and preferential attachment, respectively. The two works, along with subsequent refinements of the concepts they introduced \citep{barthelemyXprl99, newmanXprl00, krapivskyXprl00, dorogovtsevXprl00, dorogovtsevXepl00, amaralXpnas00, krapivskyXpre01, krapivskyXprl01}, spawned an impressive number of studies on networks, as evidenced by the many reviews \citep{newmanXjsp00, albertXrmp02, dorogovtsevXap02, newmanXsiam03, boccalettiXpr06, dorogovtsevXrmp08} and books \citep{dorogovtsevXbook03, pastorXbook04, newmanXbook06, barratXbook08} dedicated either specifically to this field of research or its many interdisciplinary variations. This is all the more impressive since, at least within the hard sciences, prior to the late 1990s a paper on network theory is hard to come by \citep{newmanXepl09}. For a field this young the volume of insightful findings that have accumulated until now is something to be reckoned with. For example, the structure of networks has been found crucial for their resilience to error and attack \citep{albertXnat00, cohenXprl00, callawayXprl00, cohenXprl01, pietschXpre06}, for the fast availability of information within the world-wide-web \citep{albertXnat99, pastorXprl01}, uninterrupted supply with electricity \citep{albertXpre04}, fast spread of epidemics and viral infections \citep{pastorXpre02, zanetteXpa02, barthelemyXprl04, colizzaXplos07}, robust and near flawless reproduction of organisms \citep{hartwellXnat99}, the evolution of cooperation \citep{santosXprl05, szaboXpr07, percXnjp09} and coevolution \citep{grossXjrs08, percXbs10}, the dynamics of social systems \citep{castellanoXrmp09}, and surely many other aspects of everyday life.

An interesting and potentially very revealing subset of complex networks are the social networks, of which scientific collaboration networks are a beautiful example \citep{newmanXpnas01, newmanXpre01a, newmanXpre01b, newmanXpnas04}. Notably, for a social network to be representative for what it stands -- an account of human interaction -- a consistent definition of acquaintance is important. And while it may be challenging to \textit{define} a friendship or an enemy in a consistent and precise manner \citep{moodyXajs01, moodyXasr03}, scientific collaboration is accurately documented in the final product and thus fairly straightforward to assess. Also amenable to a precise definition of connectedness are movie actors \citep{amaralXpnas00}, electric grids \citep{wattsXnat98, albertXpre04} and airports \citep{guimeraXpnas05}, for example, yet these are either approximations of social networks in that they don't really document real human contact or the level of acquaintance between people forming them is difficult to determine. As argued convincingly by \citet{newmanXpnas01}, considering scientific collaboration networks alleviates these problems to a large extend.

Here we study the evolution of a scientific collaboration network, namely that formed by Slovenia's scientists, from its very beginnings in the 1960s until the present time. Covering a time span of 50 years, the data are unique in that they provide an excellent testing ground for the ``small-world" and preferential attachment hypotheses in growing social networks. We tackle these issues similarly as outlined in previous studies on growing scientific collaboration networks \citep{newmanXpre01c, jinXpre01, ravaszXpa02, jeongXepl03, moodyXasr04}, where it has been reported, for example, that the growth is governed by linear or sublinear preferential attachment, and that as the networks grow their average degree increases while the average distance between individuals decreases. Evidences for strong clustering and models describing the growth of social networks have been presented in this context as well. Notably, for a set of different yet static scientific collaboration networks, \citet{newmanXpnas01} has shown that the average distance between different authors scales logarithmically with size. We come to results that are in agrement with these earlier observations, but for a single growing scientific collaboration network. Moreover, we show that the observed near-linear preferential attachment rate translates into the expected log-normal degree distribution fairly accurately; a detail that was previously a source of some discrepancy not just in the context of scientific collaboration networks \citep{jeongXepl03, rednerXphyst05}. In the continuation we first give information on the raw data and network construction, while the results are presented and summarized in Sections 3 and 4, respectively.

\section{Preliminaries}

Slovenia is a small country located at the heart of Europe with a population of two million.\footnote{The official Web page of Slovenia is accessible via: \href{http://www.slovenia.si/}{http://www.slovenia.si/}} It has a well-documented research history, which is made possible by SICRIS -- Slovenia's Current Research Information System\footnote{The SICRIS Web page is accessible via: \href{http://sicris.izum.si/}{http://sicris.izum.si/}} -- hosting up-to-date publication records of all Slovene scientists. At present, there are 30630 registered, including young and non-active scientists as well as laboratory personnel, which boils down the initial number to 8402 of those that are truly active research-wise or have been so in the past. By this we mean those that, to date, have at least one bibliographic unit indexed by the Web of Science. This criterium may be somewhat stringent, but it is the only one we could apply consistently. Moreover, since the publication data contain records not just of research but also of professional work and many other activities not necessarily concerning research, it is important to define a threshold for when two scientists are considered connected. Having given an interview together can hardly be compared to writing a joint research paper. In accordance with previous studies, which by default considered databases containing almost exclusively research papers and where therefore the setting of such a threshold was not necessary, we here consider two scientists as being connected if, up to the given year inclusive, they have coauthored at least one research or review paper together. A final factor slightly affecting the network size is the necessity of identifying authors on a full first and last name basis. Identification based on the last name and first initial only ($<1\%$ of Slovene population has more than one initial) has already been shown to downsize scientific collaboration networks considerably \citep{newmanXpnas01}. Here this problem is additionally amplified by the fact that certain last names are very frequent, for example because a family is traditionally involved in research over many generations, but also because in any given country certain last names are far more common than others. In our case this introduces an unacceptably large bias, grouping too many different individuals together and making them appear as a single author. On the other hand, requiring full first and last name agreement for identification practically never confuses two different authors for one, but is likely to miss out a few links if authors do not use their first names consistently. Although all publication records in SICRIS are always given with full first and last names (irrespective of whether the actual paper features just the initials of the first name), sometimes authors are evidenced in SICRIS under a given first name but then use an abbreviated (\textit{e.g.} ``Alojzij" may become ``Alojz") or internationalized (\textit{e.g.} ``Aleksander" may become ``Alex") version thereof, and SICRIS is not always consistent in eliminating such discrepancies. Challenging for parsing are also the non-ASCII characters that are quite common in Slovene names, such as {\v c}, {\'c}, {\v s} and {\v z}, but this can be solved with a high success rate by converting them all to standard ASCII counterparts via a unique-enough rule. Taking these considerations into account, we start with $30$ scientists with an average of $1.5$ collaborators in the year $1960$ and end with $7380$ scientists with an average of $10.7$ collaborators in $2010$ with a yearly time resolution in-between, assured that the error margin is highly unlikely to affect the presented results in any perceivable way. To date the $7380$ scientists forming the network in $2010$ have written $76194$ papers, altogether having $735619$ citations assigned to them. A detailed statistical analysis of individual scientific indicators, which is an interesting and vibrant subject on its own \citep{eggheXbook, hirschXpnas05, eggheXscient06, eggheXjasoc08, schreiberXnjp08, schreiberXjinform08, rousseauXjam08, schreiberXjasoc10} is, however, presented in a separate study \citep{percXjinform10}.\footnote{Tables of scientific indicators for Slovene researchers are accessible via: \href{http://www.matjazperc.com/sicris/stats.html}{http://www.matjazperc.com/sicris/stats.html}}

\section{Results}

\begin{figure}[ht!]
\begin{center}
\includegraphics[width = 16.0cm]{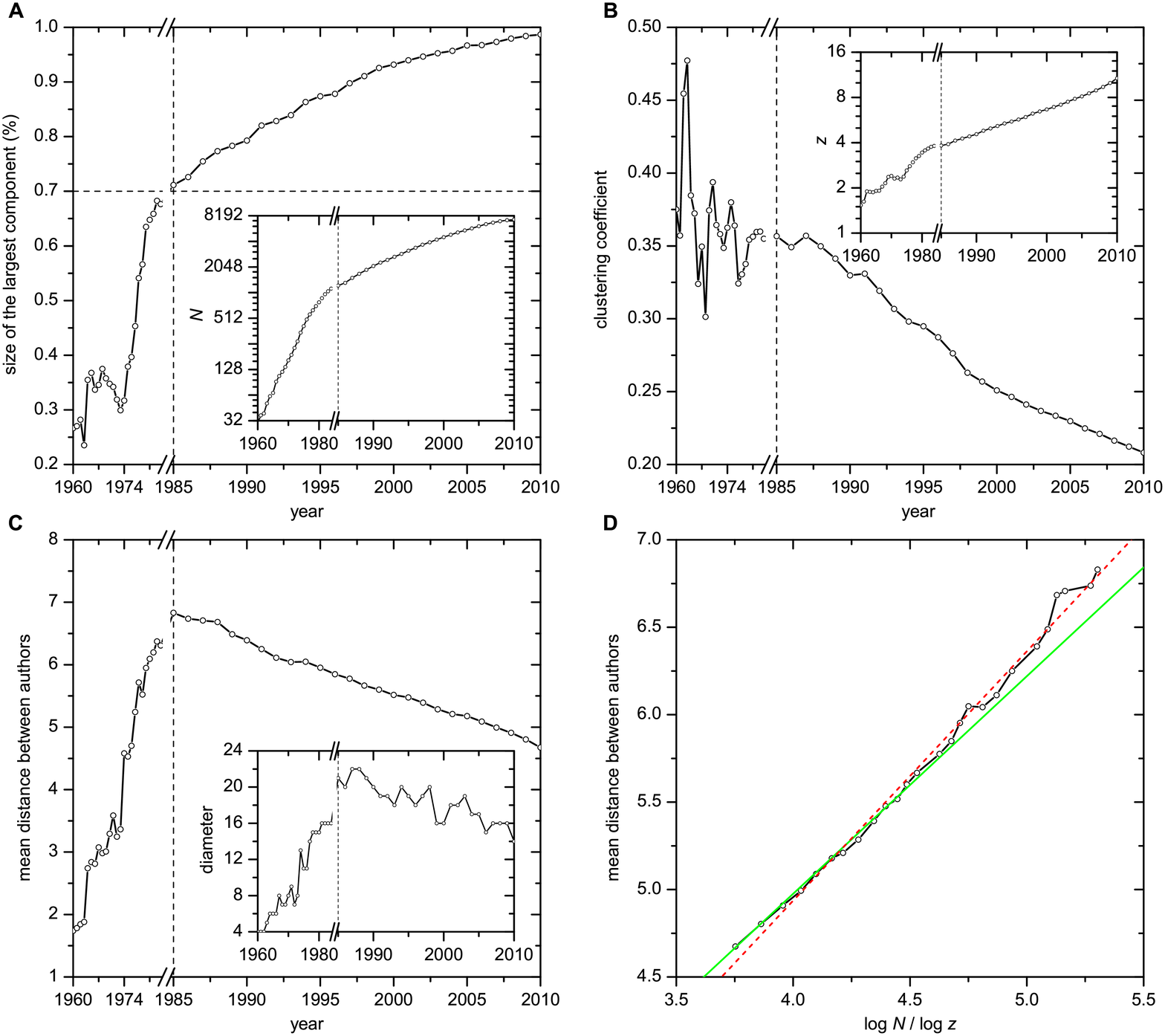}
\caption{Fundamental statistical properties of the scientific collaboration network over time. All panels have a horizontal axis break, marked by the vertical dashed line, such that the period from $1960-1984$ occupies $20\%$ ($30\%$ in insets) of the total axis span, while the period from $1985-2010$ occupies the remainder. The axis break is motivated by the emergence of the logarithmic variation of the mean distance with size (panel D) and the largest component exceeding $70\%$ of the network (panel A). This also coincides with the settling of other network properties into predictable paths past the year $1985$. (\textbf{A}) Size of the largest component in percentage of the total network size $N$. Dashed horizontal line marks the $70\%$ borderline, which is exceeded in $1985$. The inset shows network growth, starting with $N=30$ scientists in 1960 and ending with $N=7380$ in 2010. Network growth is almost exponential in time, with the curve having a slightly negative curvature on a semi-log scale. (\textbf{B}) Clustering coefficient and the average number of collaborators $z$ (inset) over time. Past the year $1985$ $z$ increases exponentially. (\textbf{C}) Mean distance between authors and the diameter (inset) over time. The tipping point is the year $1985$, following which both quantities gain a downward momentum. (\textbf{D}) Mean distance between authors past $1985$ in dependence on $\log N/\log z$. Dashed red line is the best linear fit to the data while the solid green line is the best linear fit going also through the origin.}
\label{fig1}
\end{center}
\end{figure}

We start with presenting the evolution of fundamental statistical properties of the scientific collaboration network in Fig.~\ref{fig1}. The inset of panel A shows how the network size increases with time. On a semi-log scale we could attempt making a linear fit for at least some time windows, yet the obtained exponents would not be very meaningful given that a slightly negative curvature is obviously present across the majority of the plot. We therefore satisfy ourselves with the conclusion that the network growth is almost exponential in time, which is certainly a result worth some admiration. Note that exponential growth is usually associated with online networks such as Delicious or Yahoo! Answers \citep{leskovecXkdd08}, but could hardly be expected for a scientific collaboration network. This result can be taken as evidence that science is well taken care of in Slovenia, but also that it is not yet at capacity in terms of the number of scientists that can be sustained. The main graph in panel A, on the other hand, shows the fraction of the largest component with respect to the total network size in the pertaining year. Initially the network was quite fragmented with the largest component occupying less than $40\%$ of its overall size. In good $20$ years, however, the largest component started exceeding $2/3$ of the network. Currently it occupies $98.7\%$, meaning that the vast majority of Slovenia's scientists is connected with one another via some number of intermediate acquaintances. Those not belonging to this giant component typically form small isolated groups that don't crucially affect the structure of the network. As discussed already by \citet{newmanXpnas01}, this is a desirable feature since it signalizes that science as a whole is a product of joint and strongly interdisciplinary rather than isolated efforts. This, in turn, can be seen as na important driving force behind the ever faster scientific progress and innovation.

The emergence of the giant component brings with it some noticeable changes in the way certain quantifiers of the network structure evolve over time. In order to highlight this transition, each panel of Fig.~\ref{fig1} has a horizontal axis break giving visual priority to the $1985-2010$ period, during which the giant component occupies more than $70\%$ of the network. The main graph in panel B shows that the clustering coefficient, calculated as three times the number of triangles divided by the number of connected triples of vertices [see \textit{e.g.} \citet{newmanXsiam03}], starts decreasing monotonously, while simultaneously the average number of collaborators per author (average degree of the network) assumes a steady exponential increase (see insert). Prior to $1985$ the trends are less clear and fluctuating, but qualitatively similar. Although the clustering coefficient is decreasing over the years, with around $20\%$ chance of two scientist collaborating if both have done so with a third scientist in $2010$, it can still be concluded that clustering is an inherent property of the studied scientific collaboration network. Altogether these results are in good agrement with earlier findings obtained for larger scientific collaboration networks but over a shorter time span \citep{ravaszXpa02}. Notably, large increments in the level of collaboration over an even longer time span then considered here have been reported also by \citet{grossmanXcongnum02} for a scientific collaboration network of mathematicians \citep{grossmanXcongnum95}.

Most impressive, however, is the shift in the dependence of the mean distance between authors and the diameter of the network, as depicted in panel C of Fig.~\ref{fig1} (main graph and inset). The year $1985$ constitutes a tipping point, following which the upward trend prior is replaced by a steady downward trend. However, since the diameter is obviously more prone to fluctuations than the mean distance, the tipping point in the inset is expressed less accurately, but still clearly. It is remarkable that in spite of the broad myriad of research fields covered by Slovenia's scientists, the networks still form ``small worlds" in that the path between any given pair (provided it exists) leads across only a few -- on average less than five but definitely not more than $14$ in $2010$ -- intermediate acquaintances. It is common to compare these values with what would be obtained on a random graph having the same size and average degree \citep{bollobasXbook85}. Indeed, random graphs can be considered as the simplest of models of social networks. Panel D of Fig.~\ref{fig1} shows how the average distance varies in dependence on $\log N/\log z$ (exactly the variation of the mean distance observed on random graphs) within the $1985-2010$ period, along with the best linear fit (dashed red line) and the best straight line fit going also through the origin (solid green line). Evidently, there is a considerable overlap with the logarithmic variation that can be observed for random graphs, especially past the year $1995$, where the points fall almost perfectly onto the solid green line, thus confirming the existence of the ``small-world phenomenon" in the studied scientific collaboration network. Notably, this result is interesting on its own as an empirical demonstration of logarithmic variation with size in a real growing social network.

\begin{figure}[ht!]
\begin{center}
\includegraphics[width = 15.787cm]{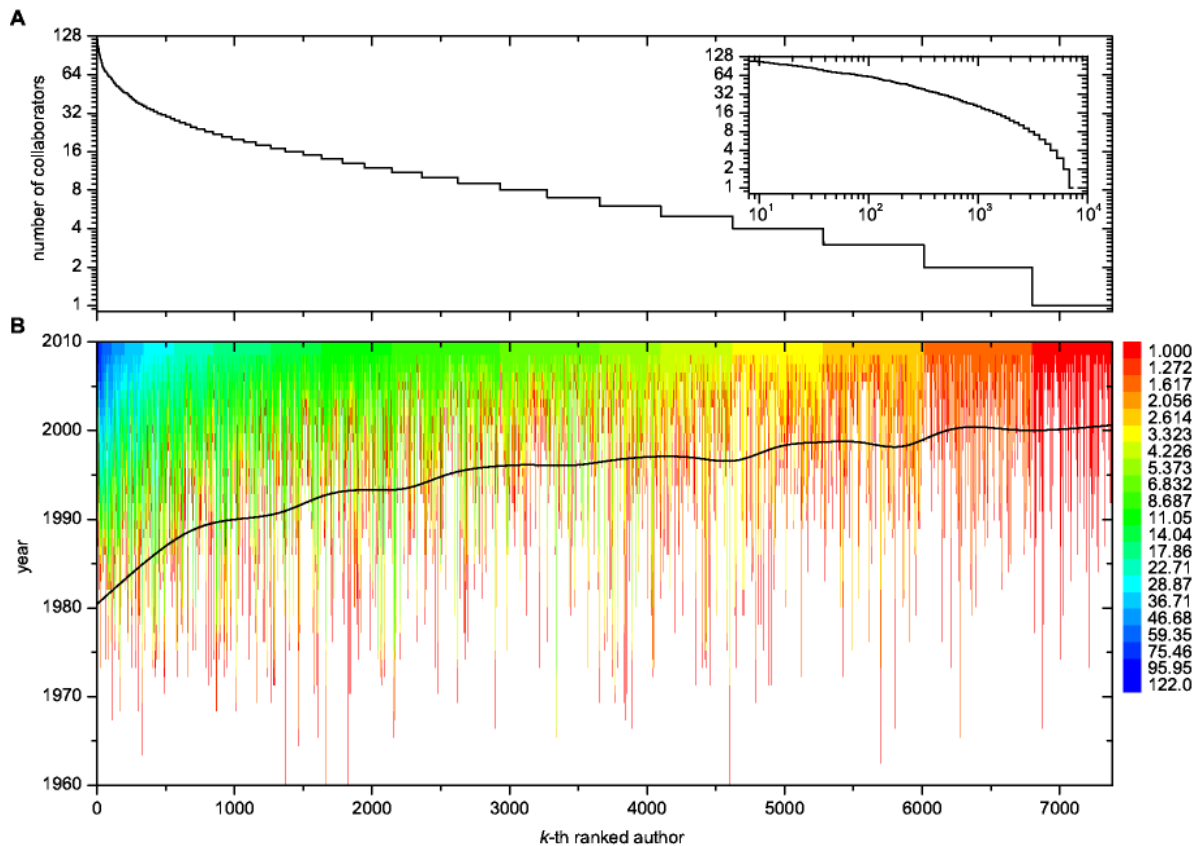}
\caption{(\textbf{A}) Zipf plot of the number of collaborators versus the $k$-th ranked author on a semi-log (inset shows log-log) scale, as obtained for the year $2010$. (\textbf{B}) Color-coded evolution of individual collaborator numbers leading to the Zipf plot in panel A. Each vertical line refers to a single author, whereby authors are ranked according to the total number of collaborators they have in the year $2010$ (see panel A). The start of each vertical line corresponds to the year the pertaining author received his/her first collaborator, \textit{i.e.} when s/he became active and thus a part of the collaboration network. The coloring denotes how the number of collaborators of each author increased over the years (from the time of becoming active till $2010$), according to the color bar on the right. Not all authors that became active early eventually gathered many collaborators, although there is a fairly clear trend towards an earlier start when approaching the $1^{\rm{st}}$ ranked author. This is evidenced by the solid black line, depicting the average starting years that were determined within non-overlapping windows each containing $300$ consecutively ranked authors.}
\label{fig2}
\end{center}
\end{figure}

In addition to the above presented results, the data enable testing the preferential attachment hypothesis in growing social networks. Growth and preferential attachment have been proposed by \citet{barabasiXsci99} as the two crucial factors for the emergence of scaling in random networks. Notably, synonymous to preferential attachment are the terms cumulative advantage \citep{priceXsci65, priceXjamsoc76} and the Matthew effect \citep{mertonXsci68}, which were proposed in the 60s but attracted considerably less attention. It is also interesting to note that the general evolutionary theory of information production processes can be applied successfully on evolving complex networks \citep{eggheXjinform07}, and that in this context the so-called ``success-breeds-success" principle \citep{eggheXjasoc95, eggheXiprocman06} was used synonymously. In essence, however, these terms all describe that the more one has (of something), the more likely it is one will eventually gain even more; and conversely, that the ones that have very little are unlikely to recover. Translating this to growing scientific collaboration networks, it means that the more collaborators one has at a given time, the more likely it is one will attract new scholars in the future. On the other hand, if one has only a few collaborators, it is less likely s/he will establish new acquaintances. Indeed, this seems like a very reasonable proposition.

Panel B of Fig.~\ref{fig2} shows how the number of collaborators increases over the years for every scientist considered in this study. Individuals are thereby enumerated according to their rank, as determined by the total number of collaborators they have in the year $2010$. Number one is the scientist that in $2010$ has the highest degree (equalling $122$), author number two has the second-largest degree (equalling $119$), and so on. The upper-most color stripe in Fig.~\ref{fig2}B (for the year 2010) thus gives the Zipf plot \citep{zipfXbook49} of the number of collaborators in the year $2010$. Since the Zipf plot can give vital clues on the expected distribution of the examined quantity, we show it in panel A of Fig.~\ref{fig2} separately on a semi-log and log-log (inset) scale. From the informetrics perspective it is noteworthy that the Lotka law, describing for example the frequency of publication by authors, is often equivalent to the Zipf law in that it also assumes an inverse square form [see \textit{e.g.} \citet{eggheXjasoc05} for an interesting treatment]. Had the Zipf plot on the log-log scale a linear outlay with the slope $\beta$, this would be equivalent to a power-law distribution of the number of collaborators $y$ of the form $P(y) \sim y^\gamma$, with $\gamma=1+1/\beta$. Furthermore, this would then imply linear preferential attachment governing the growth of the studied network. Yet this is not the case since the Zipf plot on the log-log scale has an obvious negative curvature. The semi-log scale is similarly inconclusive, thus hinting towards an interesting degree distribution and the underlying attachment rate.

Before determining this, however, it is worth examining how the year of becoming active translates to the number of collaborators one eventually acquires over the years. Although becoming active in the sense of writing the first research paper is not necessarily identical to becoming active in the sense of acquiring at least one collaborator (for the obvious reason that one can write a research paper as a sole author), this discrepancy is small enough to be negligible. The color map presented in panel B of Fig.~\ref{fig2} shows that the sooner one started acquiring collaborators, the more likely it is one has a lot of them in the year $2010$. In other words, the age of a node is roughly inversely proportional to its degree, which is characteristic for networks whose growth is governed by preferential attachment [see \textit{e.g.} \citet{bennaimXjphysa09} for a recent treatment]. However, for growing scientific collaboration networks this conclusion is obviously challenged for several reasons. Note that a sliding non-overlapping window of $300$ consecutively ranked scientists yields the average starting year as depicted by the black solid line in panel B, but individual discrepancies to this are far from uncommon. An obvious reason contributing to the deviation is that not all people are equally fond of making new acquaintances. This is true for life in general, and probably even more so for scientific collaboration. Moreover, a scientist may have gone dormant in the course of time, thus stopping the acquirement of new collaborators. Nevertheless, the statement that older scientists are more likely to have more collaborator than younger scientists is certainly reasonable and valid, and results presented in panel B of Fig.~\ref{fig2} support such a conclusion to a large extent.

\begin{figure}
\begin{center}
\includegraphics[width = 16.0cm]{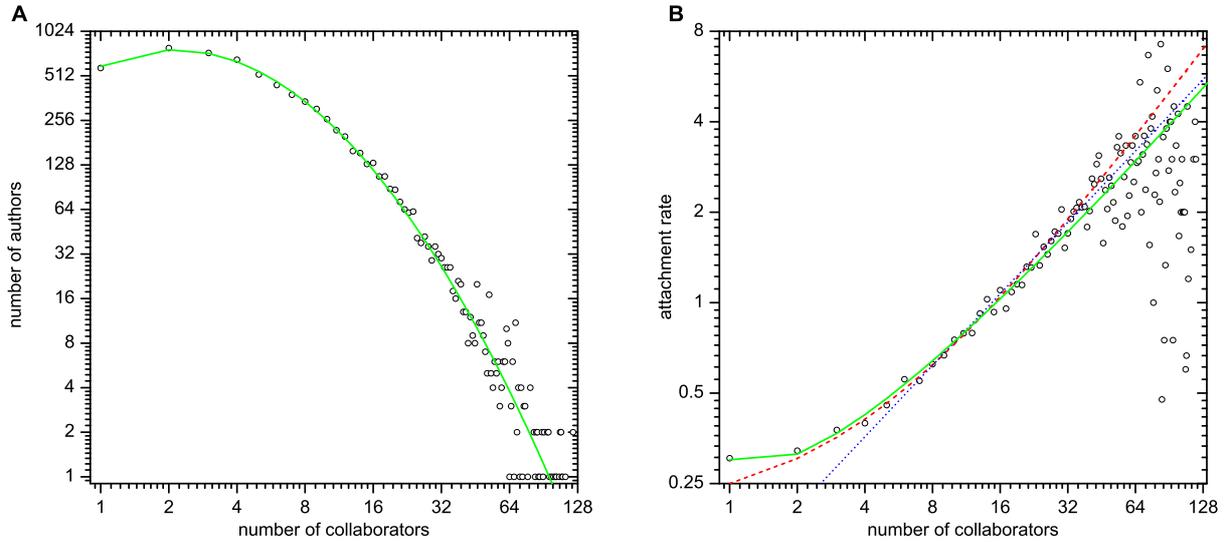}
\caption{ (\textbf{A}) Distribution of the number of collaborators $y$ on a log-log scale. Solid green line is the log-normal fit $P(y) \sim \exp[a \ln y-b(\ln y)^2]$ of the data, with $a=0.69$ and $b=0.46$. (\textbf{B}) Attachment rate $A(y)=\Delta y /y$ versus the number of collaborators on a log-log scale. Depicted is the average over 50 years, where $y$ was determined in the year $x$ and $\Delta y$ in the year $x+1$ for $x=1960, 1961, \ldots, 2009$. Solid green line is the nearly linear fit $A(y) \sim y/(1+c \ln y)$, with $c=1.32$. For comparison, dotted blue line is the sublinear fit $A(y) \sim y^g$, with $g=0.79$, and the dashed red line is the linear fit $A(y) \sim h y$, with $h=0.053$. While sublinear attachment rates give rise to stretched exponential distributions \citep{krapivskyXprl00}, log-normal distributions arise from nearly linear preferential attachment \citep{rednerXphyst05}, thus making results presented in panels A and B in good agreement with one another.}
\label{fig3}
\end{center}
\end{figure}

Turning now to testing the preferential attachment hypothesis, we first show in panel A of Fig.~\ref{fig3} the degree distribution of the network in the year $2010$, \textit{i.e.} the number of authors with a given number of collaborators. On a log-log scale the points exhibit, similarly as the Zipf plot in Fig.~\ref{fig2}, a negative curvature across the whole span of the number of collaborators. Accordingly, a power-law fit, even with an exponential cutoff, cannot describe these data accurately [see \textit{e.g.} \citet{clausetXsiam09}]. Indeed, it turns out that a log-normal form, as given in the caption of Fig.~\ref{fig3} is most fitting, as evidenced by the solid green line. While log-normal forms are typically associated with random multiplicative processes [see \textit{e.g.} \citet{rednerXpre07}], recently Redner used it for fitting the cumulative citation distribution of the Physical Review over the past $110$ years. There it was also pointed out that such distributions may arise from near-linear preferential attachment. To test this for our social network we have determined the average attachment rate $A(y)$, giving the likelihood that an author with $y$ collaborators in a given year will have $\Delta y$ more collaborators in the next year, as specified in the caption of Fig.~\ref{fig3}. This way of measuring the preferential attachment in evolving networks was proposed by \citet{jeongXepl03}, and used also for the assessment of preferential attachment in a growing scientific collaboration network, albeit over a shorter time span than considered here, by \citet{ravaszXpa02}. Results presented in the panel B of Fig.~\ref{fig3} show that the data can be fitted best by the near-linear form $A(y) \sim y/(1+c \ln y)$ (solid green line). For comparison, a linear fit (dashed red line) and a ``traditional" sublinear fit (dotted blue line) are shown as well. From this we can conclude that the growth of Slovenia's scientific collaboration network is governed by near-linear preferential attachment, which translates fairly accurately into the expected log-normal distribution of collaborators per author. Indeed, near-linear attachment rates of the form $A(y) \sim y/(1+c \ln y)$ yield log-normal distributions as depicted in panel A of Fig.~\ref{fig3} \citep{rednerXphyst05}, thus giving a coherent ending to our study.

\section{Summary}

In sum, we have studied the growth and structure of Slovenia's scientific collaboration network over the past fifty years, focusing specifically on testing the ``small-world" and preferential attachment hypotheses, but analyzing also other aspect of the network structure in detail. We have shown that there exists a tipping point in time after which the mean distance between authors and the diameter start decreasing. Accompanied by an exponential increase of the average number of collaborators per author and a near-exponential growth of the network over time, this gives rise to a logarithmic variation of the mean distance with size, much in agreement with what would be observed for growing random networks. Notably, the emergence of the tipping point coincides with the largest component exceeding $70\%$ of the network size, but roughly also with the breakup of Yugoslavia and the subsequent downfall of the socialist regime. It is unclear whether the latter fact contributed to the emergence of the transition, but the time-wise correlation gives some room for speculations. Turning back to the network growth and structure, we have shown that the clustering coefficient decreases in time, but by hoovering quite comfortably over $0.2$ in the year $2010$, it can be concluded that the clustering is still fairly strong. Finally, we have shown that the growth of the network is governed by near-linear preferential attachment that translates fairly accurately into the expected degree distribution, and that the year of becoming active scales inversely with the number of collaborators thus far acquired. Although deviations from linear preferential attachment have been reported before \citep{newmanXpre01c, ravaszXpa02, jeongXepl03}, also outside the realm of scientific collaboration networks \citep{rednerXphyst05, capocciXpre06}, an agreement between the attachment rate and the resulting distribution derived from empirical data seems to be quite rare. Altogether, however, the presented results are in agreement with previous findings, supporting the conclusion that the growth of social networks is governed by preferential attachment, and that the resulting network structure has properties characteristic for ``small worlds".

\section*{Acknowledgments}
Matja{\v z} Perc thanks Matej Horvat from Hermes SofLab for illuminating lessons on socket programming and automated information retrieval from the Internet. Financial support from the Slovenian Research Agency (grant Z1-2032) is gratefully acknowledged as well.

\end{document}